\documentclass[final]{svjour2}
\journalname{Journal of Statistical Physics}
\usepackage{amssymb,amsmath}
\usepackage[latin1]{inputenc}
\usepackage{graphicx}
\usepackage{bm}
\usepackage{color}
\usepackage{stackrel}

\newcommand{\overbar}[1]{\mkern 1.5mu\overline{\mkern-1.5mu#1\mkern-1.5mu}\mkern 1.5mu}
\newcommand{\N}{{\overbar{N}}}
\newcommand{\M}{\overbar{M}}

\renewcommand{\eqref}[1]{{\bf\ref{#1}}}

\begin{document}


\title{Scaling solution in the large population limit of the general asymmetric stochastic Luria-Delbr\"uck evolution process}

\author{David A. Kessler \and Herbert Levine}
\institute{David A. Kessler \at Department of Physics, Bar-Ilan University, Ramat-Gan, IL52900 Israel\\\email{kessler@dave.ph.biu.ac.il} \and
Herbert Levine \at 
Center for Theoretical Biological Physics, Rice University, Houston, TX 77096, USA\\\email{Herbert.Levine@rice.edu}}

\date{Received:   / Accepted: }
\maketitle

\begin{abstract}
One of the most popular models for quantitatively understanding the emergence of drug resistance both in bacterial colonies and in malignant tumors was introduced long ago by Luria and Delbr\"uck. Here, individual resistant mutants emerge randomly during the birth events of an exponentially growing sensitive population. A most interesting limit of this process occurs when the population size $N$ is large and mutation rates are low, but not necessarily small compared to $1/N$. Here we provide a scaling solution valid in this limit, making contact with the theory of Levy $\alpha$-stable distributions, in particular one discussed long ago by Landau. One consequence of this association is that moments of the distribution are highly misleading as far as characterizing typical behavior. A key insight that enables our solution is that working in the fixed population size ensemble is not the same as working in a fixed time ensemble. Some of our results have been presented previously in shortened form~\cite{pnas} \end{abstract}

\section{Introduction}

In 1943, Luria and Delbr\"uck~\cite{luria} (LD hereafter) used a simple evolutionary model together with laboratory experiments to argue that bacterial resistance  (to bacteriophage, in their specific case) arose via the selection of pre-existing mutations. Their model postulated a constant probability at which dividing normal cells would have a  resistant daughter; these mutations were assumed to not affect the growth rate until the bacteria were challenged.  The LD model and various extensions thereof have been used extensively in recent years to study the emergence of antibiotic-resistant microbes~\cite{lea,lenski} as well as chemotherapy-resistant cancer cells~\cite{coldman,moolgavkar,komarova,tomasetti,nowak}. Similar models are used to study the generation of differentiated cells from stem-like precursors~\cite{simons}.

There has been considerable effort in the mathematics and statistical physics communities devoted to understanding the probability distribution that arises from the LD process~\cite{kepler,dewanji,angerer,zheng,iwasa,kashdan}. While early efforts focused on various quasi-deterministic versions of the model where both wild-type and mutant cells or alternatively just the wild-type are assumed to undergo continuous exponential growth, more recently
interest has focussed on the fully stochastic model. A major achievement has very recently been achieved by Antal and Krapivsky~\cite{antal} who solve for the
generating function of the full joint distribution of the number of wild-types and mutants as a function of time, starting from a single wild-type individual.
Also noteworthy is the solution by Angerer~\cite{angerer} for the probability distribution of the number of mutants given a fixed total population, $N$, in the case of the symmetric  birth-only version of the stochastic model, where both wild-type and mutant have identical birth rates and cell death is excluded. However, it has proven difficult to obtain much physical insight from these solutions, and  to relate these solutions to those of the simpler quasi-deterministic models. In particular, the remarkable fact that the distribution is these simpler models  becomes~\cite{mandelbrot} a universal $\alpha$-stable (L\'evy) distribution in the limit of small mutation rate $\mu$, and large $N$ such that  $\mu N\gg 1$ has had no known analogue in the fully stochastic models. 

Here, we introduce a new strategy to solve this model, taking advantage of the fact that for most cases of interest the population size $N$ is very large and the mutation rate $\mu$ is very small. For example, a tumor may have upwards of $10^8$ cells when it is first detected and the mutation rate, even if elevated above normal human mutation rates by genomic instability, would still be $10^{-6}$ or smaller (depending on exactly what range of mutations are assumed to be able to confer resistance). Note that $\mu N$ is in general large, and hence a perturbative treatment would not be accurate. We show that if one focuses attention on the fixed $N$ ensemble, one can derive an alternative generating function equation which leads immediately to a simple integral expression for the probability distribution. One can recover the asymptotic properties of this distribution for the general case where the resistant population may have a different fitness than the wild-type. 

The outline of this work is as follows. To illustrate our methodology and to lay out the general features of the LD process, in particular  that it gives rise to a distribution $P(m)$ for the
number $m$ of mutants which is heavy-tailed, we first present the special case of zero death rate. We show the general features of the quasi-deterministic model, including the fact that at  large $\mu N$ one obtains a universal $\alpha$-stable distribution
 is actually valid for the fully stochastic version as well. Next, we consider cell death in addition to division and mutation, and show that even for this case we recover for large $\mu N$  the limiting Landau distribution (for equal birth rates) and its generalization for the asymmetric case. Finally, we compare our results to those of Antal and Krapivsky, who formally solve the same problem  with a fixed time ensemble. Some of our results have appeared previously in abbreviated form~\cite{pnas}.

\section{The symmetric pure birth model}

To illustrate our basic ideas, we start with the relatively simple case of no cell death (i.e. a pure birth/mutation process) and with equal birth rates for both the wild-type and mutant subpopulations. If we start from a fixed number $N_0$ of initial wild-type cells  and wish to study the ensemble at fixed total population $N$,  the number of events is just $N-N_0$. Rather than computing the probability function at fixed time, we instead use the event number as the independent variable. Then, $P_N(0)$ is just the probability that in the $N-N_0$ births since the beginning, there were no mutations, thus
\begin{equation}
P_N(0) = (1-\mu)^{N-N_0} .
\label{PN0}
\end{equation}
The master equation for $N>N_0$, $m>0$ reads
\begin{align}
P_{N+1}(m) &= \frac{1}{N}\Big[ P_N(m-1)\mu(N-m+1) + P_N(m) (1-\mu)(N-m) \nonumber\\
&\qquad\qquad {} + P_N(m-1)(m-1)\Big]  .
\end{align}

To derive what we will refer to as the scaling solution for the probability distribution, we will keep the leading order in $N$ term for each order in $\mu$. Later, we will see exactly when this is valid. It is easy to check that this involves keeping the coefficient of $N^k$ for terms proportional to $\mu ^k$. Let us denote this coefficient in $P_N (m)$ as $A^k_m$, so that
\begin{equation}
P_N(m) \approx \sum_{k=0}^\infty A^k_m y^k;   \qquad  y\equiv \mu N.
\label{expan}
\end{equation}. 
Substituting this term into the master equation, we can check that the leading order in $N$ condition (the $(\mu N)^k$ term is automatically satisfied, but matching the next order in $N$ term (proportional to $\mu ^k N^{k-1}$) leads to the non-trivial condition for $m,k\ge1$:
\begin{equation}
(k+m)  A^k_m = A^{k-1}_{m-1}- A^{k-1}_m+ (m-1)A^k_{m-1}  ,\label{scaling-master}
\end{equation}
with
\begin{equation}
A_0^k = (-1)^k/k!; \qquad \qquad A_m^0=0,\quad  (m=1\ldots\infty) .
\end{equation}
The former condition comes from our exact solution of $P_N(0)$, Eq. (\ref{PN0}), above and the latter from the fact that $P_N(m)$ for all $m>0$ vanishes if $\mu=0$.
To solve this equation, we define the generating function 
\begin{equation}
F(x,y) \equiv \sum _{k=1,m=1} A^k_m x^m y^k,
\end{equation}
in terms of which the generating function, $P_N(x)$, of the probability distribution is given by
\begin{equation}
P_N(x) = F(x,\mu N) + e^{-y} .
\end{equation}
From the recursion relation Eq. (\ref{scaling-master}) we easily obtain the
differential equation
\begin{equation}
\left[y\frac{\partial}{\partial y} + x(1-x)\frac{\partial}{\partial x}\right]F(x,y) = -y(1-x)F + xye^{-y} .
\end{equation}

As already discussed in \cite{pnas}, we can solve this equation by first converting it to a homogeneous equation via the transformation $F=G - e^{-y}$, so that 
\begin{equation}
\left[y\frac{\partial}{\partial y} + x(1-x)\frac{\partial}{\partial x}\right]G(x,y) = -y(1-x)G m ,
\end{equation}
with $G(y=0) = 1$. 
The method of characteristics tells us to write $G(x,y)=G(x(y),y)$, choosing $dx/dy = x(1-x)/y$, so that the characteristic is
\begin{equation}
x=\frac{y}{\alpha + y} ,
\end{equation}
where $\alpha$ labels the characteristic.  Then the equation reads
\begin{equation}
y\frac{dG}{dy} = - \frac{\alpha y}{\alpha+y}G ,
\end{equation}
with the solution 
\begin{equation}
G = \left(1 + \frac{y}{\alpha}\right)^{-\alpha}; \qquad\qquad \;\;\; 
\alpha = \frac{y(1-x)}{x} ,
\end{equation}
so that
\begin{equation}
F = (1-x)^{y(1-x)/x} - e^{-y} , \label{lea-coulson}
\end{equation}
and thus
\begin{equation}
P_N(x) = (1-x)^{\mu N(1-x)/x} .
\label{LC}
\end{equation}

This result was derived initially by Lea and Coulson \cite{lea}, for a model which neglected the stochastic nature of the birth-death dynamics for the wild-type population. In our derivation, going to the scaling limit of $N$ large, $m\ll N$, has resulted in a distribution which has no $N_0$ dependence. Since when $N_0$ is large, the wild-type stochasticity is negligible, the fully stochastic treatment has to reduce to the Lea-Coulson result in the scaling limit.

\begin{figure}
\includegraphics[width=0.7\textwidth]{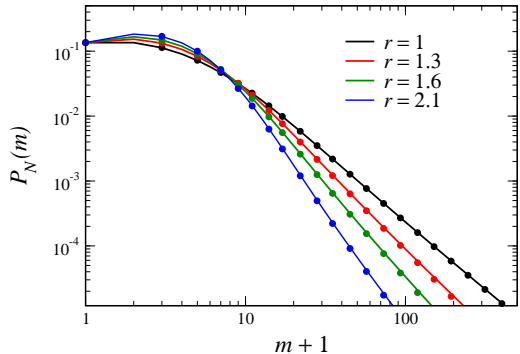}
\caption{The small $\mu$, large $N$ approximate distribution for $\mu N=2$ for the pure birth-model, with the ratio of wild-type to mutant birth rate $r=1$, $1.3$, $1.6$ and $2.1$ (solid lines).  This is shown together with the results (filled circles) of a direct numerical solution of the master equation for the parameters $\mu=0.004$, $N=500$.}
\label{LCasym}
\end{figure}

One can numerically evaluate  the Lea-Coulson distribution from the generating function via a contour integration to pick out $P_N(m)$:
\begin{equation}
P_N(m) = \frac{1}{2\pi i} \oint \frac{dz}{z^{m+1}} (1-z)^{\mu N(1-z)/z} .
\end{equation}
This is more conveniently computed by moving the contour to hug the branch cut along the real axis $x>1$:
\begin{equation}
P_N(m) \approx \frac{1}{\pi} \int_1^\infty dx\, x^{-m-1} (x-1)^{-\mu N(x-1)/x} \sin \frac{\pi \mu N (x-1)}{x} .
\end{equation}
A computation of this for the case $\mu N=2$ is shown in Fig. \ref{LCasym}, (labelled $r=1)$, together with results of a direct numerical solution of the master equation for the parameters
$\mu=0.004$, $N=500$.  The agreement is excellent even for this not so tiny value of $\mu$.

It is clear by looking at the small $x$ behavior that Eq. (\ref{LC}) correctly reproduces $P_N(0)\approx e^{-\mu N}$. In the small $\mu N$ limit, 
\begin{equation}
P_N(x) \approx 1 + \mu N \frac{1-x}{x}\ln(1-x) = 1 - \mu N \frac{1-x}{x} \sum_{m=1}^\infty  \frac{x^m}{m} ,
\end{equation}
so that for $m\ge 1$
\begin{equation}
P_N(m) \approx  \mu N \left(\frac{1}{m} - \frac{1}{m+1}\right) =\frac{ \mu N }{m(m+1)} ,
\end{equation}
which is monotonically decreasing and has a fat-tailed $1/m^2$ decay.  This $1/m^2$ behavior at large $m$ is in fact true for all $\mu N$, and is a direct result of the logarithmic singularity of $P_N(x)$ at $x=1$. 

The large $\mu N$ limit is also tractable, for $m$'s of order $\mu N$.
We write $m=\mu N \widetilde m$, and write $x=1 + t/(\mu N)$, giving
\begin{equation}
P_N(m) = \frac{1}{\mu N \pi} \int_0^\infty \!\!dt\, e^{-\tilde{m}t -t\ln t/y}\sin \pi t = \frac{1}{\mu N \pi} \int_0^\infty\!\! dt\, e^{-(\tilde{m}-\ln y)t -t\ln t}\sin \pi t .
\end{equation}
This is just a Landau distribution, the one-sided $\alpha$-stable (L\'evy) distribution with index $\alpha=1$, written in terms of the original variables as $P_\textrm{Landau}(m/\mu N - \ln \mu N)$ times the normalization factor $1/\mu N$. The Landau distribution rises very rapidly from zero, reaches it maximum and then falls with a  $1/m^2$ fat tail.  The appearance of the $\alpha$-stable distribution in the quasi-deterministic model~\cite{mandelbrot,pnas} is a consequence of the fact that the total number of mutants is the sum of the mutants descended from each mutation event, and from large $N$ these are
independent and identically distributed with a power-law tail, and thus the conditions of the generalized central-limit theorem are satisfied. The Landau distribution arises in the fully stochastic model for the same reasons; the only difference being an overall shift in the distribution.

\begin{figure}
\includegraphics[width=0.7\textwidth]{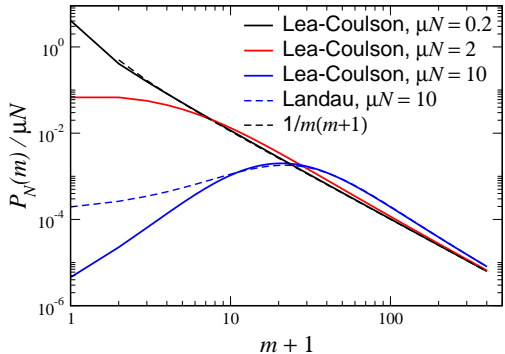}
\caption{
The probability distributions $P_N(m)/\mu N$ as given by the Lea-Coulson formula for $\mu N = 0.2$, $2$ and $10$.  In addition, the small $\mu N$
limiting form $1/m(m+1)$ and the large $\mu N$ Landau distribution are also shown.  It is seen that the Landau distribution correctly describes the distribution except for $m$'s of order unity.}
\label{figLC}
\end{figure}

Thus, the Lea-Coulson distribution is seen to interpolate between the monotonic $\mu N/m(m+1)$ distribution at small $\mu N$ to the fat-tailed one-sided
$\alpha$-stable distribution with its sharp rise for small $m$ and $1/m^2$ falloff at large $m$.  A signpost value is $\mu N=2$, above which $P_N(m)$ is no longer monotonic.  The position of the maximum moves rightward with increasing $\mu N$, asymptotically behaving as $m_{\textit{max}} \approx \mu N(\ln\mu N - 0.22)$.  This analysis is summarized in Fig. \ref{figLC}.

For $m$ of order $N$, the Lea-Coulson result breaks down because we have neglected terms that become of comparable order in this new regime. This outer regime~\cite{pnas} describes how the distribution must drop to zero as we approach $m= N-N_0$; after all, the most extreme possibility is that every division resulted in adding one more mutant. Since $\mu$ is small, this part of the distribution arises when one of the first births lead to a mutant and hence the mutant sub-population has as much chance to expand exponentially as does the wild-type. To describe this scenario, one need only keep terms linear in $\mu$. This can be done explicitly in the symmetric birth-only model~\cite{pnas}.  However, even this problem is difficult to solve exactly in the general case, and is only of 
mathematical interest, since this outer part of the distribution describes extremely rare events and have essentially no physical relevance.

\section{The non-symmetric pure birth problem}

We now extend the previous results to account for a fitness (aka birth rate) difference between the wild-type and the resistant subpopulations. Normally it would be exacted that alleles conferring resistance are less fit in the absence of any drug, and hence we expect  the division rate ratio $r$ of wild-type to resistant cell to be greater than one. The master equation connecting the mutant number distribution at one overall population size to the next overall population size, must take into account the relative probability that it is a wild-type cell that divides versus that it is a mutant cell which divides. If the current population has $m$ mutant cells and hence $N-m$ wild-type ones, this relative probability is 
$(N-m)r/((N-m)r +m)$.
Then the equation clearly becomes
\begin{align}
P_{N+1}(m) &= (1-\mu)P_N(m) \frac{(N-m)r}{(N-m)r + m} \nonumber\\
&\qquad{}+ \frac{P_N(m-1)}{(N-m+1)r + (m-1)} \left[ (N-m+1)r\mu 
 + (m-1)\right] .
\end{align}

Again, we solve this equation in the limit of large $N$, small $\mu$ by keeping only the leading term in $N$ for each poet of $\mu$, again using the expansion Eq. (\ref{expan}).  Substituting this into the master equation, we obtain a generalization of Eq. (\ref{scaling-master}):
\begin{equation}
(k+\frac{m}{r})  A^k_m = A^{k-1}_{m-1}- A^{k-1}_m+ \frac{(m-1)}{r}A^k_{m-1} . \label{scaling-master-r}
\end{equation}
Following the same procedure as before, we transform this into an equation for the generating function, 
\begin{equation}
\left[y\frac{\partial}{\partial y} + \frac{x(1-x)}{r}\frac{\partial}{\partial x}\right]F(x,y) = -y(1-x)F(x,y) + xye^{-y} .
\end{equation}
Again, the inhomogeneous solution is $-e^{y}$, and the homogeneous solution satisfies
\begin{equation}
\left[y\frac{\partial}{\partial y} + \frac{x(1-x)}{r}\frac{\partial}{\partial x}\right]G(x,y) = -y(1-x)G(x,y) ,
\end{equation}
with $G(y=0)=1$. Again, we solve by the method of characteristics, with the characteristic labelled by $\alpha$ given by
\begin{equation}
x_\alpha(y) = \frac{y^{1/r}}{\alpha + y^{1/r}} ,
\end{equation}
so that
\begin{equation}
\frac{d}{dy}G(y,x(y)) = -\frac{\alpha}{\alpha + y^{1/r}}G,
\end{equation}
with the solution
\begin{equation}
G = \exp\left(-\int_0^y \frac{\alpha}{\alpha + t^{1/r}} dt\right) = \exp\left(-y\,{}_2F_1\!\left(1,r;1+r;-\frac{y^{1/r}}{\alpha}\right)\right),
\end{equation}
where ${}_2F_1$ is a hypergeometric function.
Substituting $\alpha = y^{1/r}(1-x)/x$, we get
\begin{equation}
G  = \exp\left(-y\,{}_2F_1\!\left(1,r;1+r;-\frac{x}{1-x}\right)\right)=\exp\left(-y(1-x)\,{}_2F_1(1,1;1+r;x)\right) ,
\end{equation}
where we have employed a linear transformation identity, so that finally
\begin{equation}
P_N(x) = \exp\left(-\mu N(1-x) \,{}_2F_1\!\left(1,1;1+r;x\right)\right) .
\end{equation}

Again, to evaluate $P_m$ we integrate around the branch cut, which again extends rightward from $x=1$.  Using standard identities involving hypergeometric functions, we have
\begin{align}
{}_2F_1\!\left(1,r;1+r;-\frac{x}{1-x}\right)
 &= \frac{r}{r-1}\left(\frac{1-x}{x}\right){}_2F_1\!\left(1,1-r;2-r;-\frac{1-x}{x}\right) \nonumber\\
&\qquad  {} + \frac{\pi r}{\sin \pi r}\left(\frac{1-x}{x}\right)^r{}_2F_1\!\left(r,0;r;-\frac{1-x}{x}\right) \nonumber\\
&= \frac{r}{r-1}\left(\frac{1-x}{x}\right){}_2F_1\!\left(1,1-r;2-r;-\frac{1-x}{x}\right) \nonumber\\
&\qquad {} + \frac{\pi r}{\sin\pi r}\left(\frac{1-x}{x}\right)^r .
\label{Fr}
\end{align}
This allows us to evaluate the discontinuity along the branch cut, giving
\begin{align}
P_N(m) &= \frac{1}{\pi}\int_1^\infty \frac{dx}{x^{m+1}} \exp\left(\mu N\frac{r(x-1)}{(r-1)x}\,{}_2F_1\left(1,1-r,2-r,\frac{x-1}{x}\right)\right)\nonumber\\
&\qquad\qquad \times\exp\left(-\mu N\frac{\pi r}{\tan\pi r}\left(\frac{x-1}{x}\right)^r\right)
 \sin\left(\mu N\pi r\left(\frac{x-1}{x}\right)^r\right) .
\end{align}
Again, Fig. \ref{LCasym} presents a comparison of this expression with results of a direct numerical solution of the master equation for the parameters
$\mu=0.004$, $N=500$.  The agreement is again excellent.

The non-analyticity at $x=1$ as before governs the large $m$ asymptotics (subject to $m \ll N$, as usual). To investigate the behavior near $x=1$, we use Eq. (\ref{Fr}) to obtain ($t\equiv 1-x$)
\begin{equation}
{}_2F_1\!\left(1,r;1+r;-\frac{x}{1-x}\right) \approx \frac{r}{r-1} t + \frac{r\pi}{\sin \pi r} t^r .
\end{equation}
By standard Tauberian theorems, this corresponds to a $m^{1+r}$ decay at large $m$, independent at $\mu N$.  In particular, for small $\mu N$, we have
\begin{align}
P_N(x) &\approx 1 - \mu N  (1-x)\,{}_2F_1\!\left(1,1;1+r;x\right) \nonumber\\
&= 1 - \mu N + \mu N \sum_{m=1}^\infty \frac{(m-1)!\Gamma(1+r)r}{\Gamma(r+m+1)} ,
\label{smallmur}
\end{align}
so that for $m\ge 1$:
\begin{equation}
P_N(m) \approx \mu N  \frac{(m-1)!\Gamma(1+r)r}{\Gamma(r+m+1)} \stackrel[\scriptscriptstyle m\to \infty]{\displaystyle \to}{} \mu N \frac{r\Gamma(1+r)}{m^{1+r}} .
\end{equation}

For large $\mu N$, as we saw in the case of $r=1$, the contour integral giving $P_N(m)$ is dominated by the $x\approx 1$ singularity of the integrand.
Thus,
\begin{equation}
P_N(m) \approx \int_0^\infty  dt \, e^{-mt} \exp\left(-\mu N\left[ \frac{r}{r-1} t + \frac{r\pi}{\sin \pi r} t^r\right]\right) .
\label{PNmr}
\end{equation}
This is nothing but the one-sided $\alpha$-stable distribution, now with index $\alpha=r$, consistent with the power-law decay. To see this, note that
the characteristic function, $\phi$ (the Fourier transform of the probability density) of the $\alpha$-stable distribution with index $r$ has the form
\begin{equation}
\phi =\exp\left[ it\nu - |ct|^r(1-i \textrm{sgn}(t)\tan(\pi r/2))\right] ,
\end{equation}
where $\nu$ is called the shift parameter, which is just the mean of the distribution, and $c$ is the scale parameter.  By rotating  $t \to -it$ in Eq. (\ref{PNmr}), we get for the shift parameter
\begin{equation}
\nu = \mu N \frac{r}{r-1} ,
\end{equation}
which is indeed the mean of the  distribution Eq. (\ref{smallmur}).  For the scale parameter, we get
\begin{equation}
c = \left[ \mu N \frac{r \pi}{\sin \pi r} \cos \frac{\pi r}{2} \right]^{1/r} = \left[ \mu N \frac{r \pi}{2\sin \pi r/2}  \right]^{1/r} .
\end{equation}

\section{General Asymmetric process with death}

We now study the general asymmetric LD model with death, so that each event is either a birth event with rates  $b_w$, $b_m$ or a death event with rates $d_w$, $d_m$.   
We start from the probability distribution as a function of  time. The master equations for $P (N,m,t)$, where $N$ denotes the total (w.t. + mutated) population and $m$ is the number of mutants, are:
\begin{align}
\dot{P}(0,0) &= d_w P(1,0) + d_b P(1,1)  ,\nonumber \\
\dot{P}(N,m) &= b_w (N-1-m)(1-\mu) P(N-1,m) + b_w \mu (N-m) P(N-1,m-1) \nonumber\\
&\quad{}+ b_m (m-1)P(N-1,m-1)\nonumber\\
&\quad {}+ d_w (N+1-m)P(N+1,m) + d_m (m+1) P(N+1,m+1)\nonumber\\
&\quad {}-\left[(b_w+d_w)(N-m)+(b_m +d_m)m\right]P(N,m); \qquad\qquad (N>1),
\end{align}
with $P(N_0,0)=1$ at $t=0$.
 
To simplify these expressions, we make use of the fact that we are interested in a fixed value of $N$, independent of how much time has elapsed since the
beginning. From the biology perspective, what one knows (ideally) is the size of the tumor at discovery, with no idea how long the tumor has been growing.
Thus, we are interested in the conditional probabilities 
\begin{equation}
P(m|N) \equiv (Z_N)^{-1} \int_0^\infty P(N,m;t) dt \equiv  (Z_N)^{-1} T_N(m) ;\qquad Z_N  \equiv \sum_m T_N(m) .
\end{equation}
Integrating  the master equation over time  leads to the following equation for $T_N(m), N \ge 1$:
\begin{align}
\Big[(b_w+d_w)(N-m)&+(b_m+d_m)m\Big]T_N(m) - \delta_{N,N_0}\delta_{m,0} \nonumber\\
&=  b_w (N-1-m)(1-\mu) T_{N-1}(m)\nonumber\\
&\quad{} + b_w \mu (N-m) T_{N-1}(m-1) + b_m (m-1)T_{N-1}(m-1)\nonumber\\
&\quad {}+ d_w (N+1-m)T_{N+1}(m) + d_m (m+1) T_{N+1}(m+1) .
\label{Teq}
\end{align}
with the definition $T_0(m)\equiv 0$. Again we assume $T_N(m)=\sum_k (\mu N)^k A_m^k$.  Then, plugging this into the recursion relation, and noting  that as always the $N^k$ term cancels, we find the generalized scaling-limit recursion
\begin{align}
A_m^k \left[m(b_m+d_m) +k(b_w-d_w) \right] &= b_w A_{m-1}^{k-1} -  b_w A_m^{k-1} + b_m(m-1)A_{m-1}^k  \nonumber\\
&\qquad{}+ d_m(m+1)A_{m+1}^k .
\end{align}

By now, the procedure for solving this equation should be obvious. We rewrite this systems as an equation for the generating function'
 \begin{equation}
\left[(b_w-d_w) y \frac{\partial}{\partial y} + (x(b_m+d_m) - b_mx^2 - d_m)\frac{\partial}{\partial x}\right]F(x,y) = -b_wy(1-x)F + {\cal{I}} ,
\end{equation}
where ${\cal{I}}$ is the inhomogeneous term.  Again $x=1$ is the singular point with the coefficient of $\partial F/\partial x$ vanishes. Defining $r\equiv
(b_w-d_w)/(b_m-d_m)$, the ratio of {\emph{net}} birth rates, the characteristic is given by
\begin{equation}
x(y)=\frac{y^{1/r} + \alpha d_m}{y^{1/r} + \alpha b_m} .
\end{equation}
The homogeneous equation now reads
\begin{equation}
(b_w-d_w) y \frac{dG}{dy} = -b_w y\frac{\alpha (b_m-d_m)}{y^{1/r}+\alpha b_m} G .
\end{equation}
Notice that except for factors, this is the exact same equation we encountered in the asymmetric pure-birth case, and so the solution is
immediate:
\begin{align}
G &={\cal{N}}  \exp\left(-\frac{b_w}{rb_m}y\,{}_2F_1\!\left(1,r;1+r;-\frac{y^{1/r}}{\alpha}\right)\right) \nonumber\\
&= {\cal{N}}  \exp\left(-y\frac{b_w}{rb_m}\,{}_2F_1\!\left(1,r;1+r;-\frac{b_mx-d_m}{b_m(1-x)}\right)\right) ,
\end{align}
where ${\cal{N}}$ is a normalization prefactor.  Our desired conditional probability generating function, which we denote by $P_N(x)$ is then
\begin{equation}
P_N(x) = \exp\left(-\mu N\frac{b_w}{rb_m}  \,{}_2F_1\!\left(1,r;1+r;-\frac{b_m x - d_m}{b_m(1-x)}\right)\right) .
\end{equation}
We see that in the pure birth case, $d_w=d_m=0$, this reproduces the results of the previous section.

The large $m$ behavior is, as we have seen, determined by the behavior in the vicinity of $x=1$.  Writing $x=1-t$, we have
\begin{equation}
{}_2F_1\!\left(1,r;1+r;-\frac{b_m x - d_m}{b_m(1-x)}\right) 
\approx \frac{rb_m}{(r-1)(b_m-d_m)} t + \frac{r\pi}{\sin \pi r} \left(\frac{b_m}{b_m-d_m}t\right)^r .
\label{gensmallt}
\end{equation}
The analytic structure is identical to the asymmetric pure-birth case, and so $P(m|N)$ again decays as $m^{1+r}$:
\begin{equation}
P(m|N) \approx\mu N \left(\frac{b_m}{b_m-d_m}\right)^r\frac{b_w \Gamma(1+r)}{b_m m^{1+r}} .
\label{BDbigm}
\end{equation}
Though the exponent is only a function of the
the ratio of the {\em{net}} birth rates,  the coefficient is more complicated.  

\begin{figure}
\includegraphics[width=0.7\textwidth]{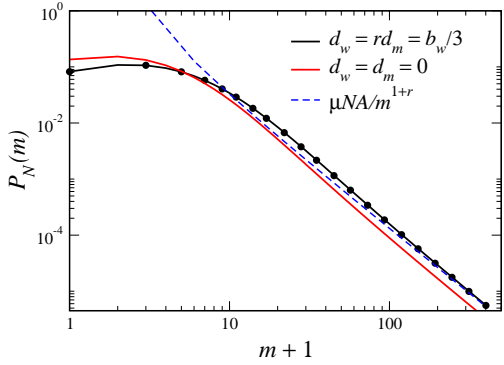}
\caption{The generalized Lea-Coulson distribution $P(m|N)$ for the asymmetric birth-death model, together with a direct solution of the master equation, Eq. (\ref{Teq}) for $\mu=0.004$, $N=500$.  Here $b_w=3/4$, $d_w=1/4$, $b_m = b_w/r$, $d_m = d_w/r$, with $r=1.3$. The asymptotic power law, Eq. (\ref{BDbigm}) is also shown. For comparison, the pure birth distribution with the same $r=1.3$ is shown.  }
\label{BDfig}
\end{figure}

We can again evaluate $P(m|N)$ from our generating function via an integration along the branch cut:
\begin{align}
P_N(m) &= \frac{1}{\pi}\int_1^\infty \frac{dx}{x^{m+1}} \exp\left(\mu N\frac{b_w(x-1)}{(r-1)(b_mx-d)}\,{}_2F_1\left(1,1-r,2-r,\frac{b_m(x-1)}{b_mx-d}\right)\right)\nonumber\\
&\quad\times\exp\left(-\mu N\frac{ \pi b_w }{b_m \tan\pi r}\left(\frac{b_m(x-1)}{b_mx-d}\right)^r\right) \sin\left(\mu N\frac{\pi b_w}{b_m} \left(\frac{b_m(x-1)}{b_mx-d}\right)^r\right) .
\end{align}
We present in Fig. \ref{BDfig} such a calculation for the case $b_w=3/4$, $d_w=1/4$, $b_m = b_w/r$, $d_m = d_w/r$, with $r=1.3$.  Also shown is a direct numerical solution of the master equation, Eq. (\ref{Teq}), for $\mu=0.004$, $N=500$, which confirms our asymptotic analysis.  For comparison, the pure birth distribution with the same $r=1.3$ is shown.

 For small $\mu N$, we can write $P(m|N)$ explicitly, since
 \begin{align}
 P_N(x) &\approx 1 - \mu N \frac{b_w}{rb_m}  \,{}_2F_1\!\left(1,r;1+r;-\frac{b_m x - d_m}{b_m(1-x)}\right) \nonumber\\
 &= 1 -  \mu N \frac{b_w}{rb_m}\sum_{k=0}^\infty \frac{1}{k!}\left(\frac{d_m}{b_m(1-x)}\right)^k \frac{\Gamma(1+k)\Gamma(r+k)\Gamma(1+r)}{\Gamma(1)\Gamma(r)\Gamma(1+r+k)}\nonumber\\
 &\qquad\qquad\qquad\qquad \times \,{}_2F_1\!\left(1+k,r+k;1+r+k;-\frac{x}{1-x}\right) \nonumber\\
 &=1 -  \mu N \frac{b_w}{rb_m}\sum_{k=0}^\infty \left(\frac{d_m}{b_m}\right)^k \frac{r}{r+k} (1-x)\,{}_2F_1\!\left(1,1+k;1+r+k;x\right) \nonumber\\
 &=1 -  \mu N \frac{b_w}{b_m}\sum_{k=0}^\infty \left(\frac{d_m}{b_m}\right)^k \frac{1-x}{r+k} 
 \sum_{m=0}^\infty \frac{\Gamma(1+m)\Gamma(1+k+m)\Gamma(1+r+k)}{\Gamma(1)\Gamma(1+k)\Gamma(1+r+k+m)
 m!}x^m\nonumber\\
 &=1 +  \mu N \frac{b_w}{b_m}\sum_{k=0}^\infty \left(\frac{d_m}{b_m}\right)^k\left[ -\frac{1}{r+k} +
 \sum_{m=1}^\infty \frac{r\Gamma(k+m)\Gamma(r+k)}{\Gamma(1+k)\Gamma(1+r+k+m)}x^m\right]\nonumber\\
 &=1 + \mu N \frac{b_w}{rb_m}\Big[-{}_2F_1\left(1,r;1+r;\frac{d_m}{b_m}\right) \nonumber\\
 &\qquad\qquad\qquad\qquad{}+ \sum_{m=1}^\infty\frac{r^2\Gamma(r)\Gamma(m)}{\Gamma(r+m+1)}\,{}_2F_1\left(r,m;r+m+1;\frac{d_m}{b_m}\right) x^m\Big] ,
 \end{align}
 so that, for $m>0$,
 \begin{equation}
 P(m|N)\approx \mu Nr \frac{b_w}{b_m}\frac{\Gamma(r)\Gamma(m)}{\Gamma(r+m+1)}\,{}_2F_1\left(r,m;r+m+1;\frac{d_m}{b_m}\right)\equiv p(m|N) ,
 \end{equation}
 and 
 \begin{equation}
 P(0|N) \approx  1 - \mu N \frac{b_w}{rb_m}  \,{}_2F_1\!\left(1,r;1+r;\frac{ d_m}{b_m}\right) \equiv 1 + p(0|N) .
 \end{equation}
 It is straightforward, though tedious, to show that 
 \begin{equation}
 t_N(m) = \frac{1}{b_w N } p(m|N)
 \end{equation}
 is an {\em{exact}} solution  of the small $\mu$ version of Eq. (\ref{Teq}) (using the fact that to zeroth order in $\mu$, $T_N(0) \approx 1/(b_w N)$:
 \begin{align}
 [(b_w + d_w)(N-m) &+ (b_m + d_m)m]t_N(m) \nonumber\\
 &= b_w (N-1-m) t_{N-1}(m)  + b_m (m-1)t_{N-1}(m-1)\nonumber\\
&\quad{} 
+ d_w (N+1-m)t_{N+1}(m) + d_m (m+1) t_{N+1}(m+1) ,\nonumber \\
 [(b_w + d_w)(N-1) &+ (b_m + d_m)]t_N(1) \nonumber\\
 &= b_w (N-2) t_{N-1}(1)  + d_w Nt_{N+1}(1) + 2 d_m  t_{N+1}(2) + \mu , \nonumber\\
 [(b_w + d_w)N]t_N(0) 
 &= b_w (N-1) t_{N-1}(0)+ d_w (N+1)t_{N+1}(0) \nonumber\\
 &\qquad{}+ d_m  t_{N+1}(1) - \mu ,
\end{align}
with $0\le m \le \infty$.  However, the finite $N$ system is truncated, with the equation for $t_N(N)$ being different, so that this solution does not represent an exact solution of this truncated system. Nevertheless, it turns out that $t_N(0)$ and $t_N(N)$ decouple from the other equations, and in fact
$T_N(m) =  t_N(m)$ {\em{is}} the exact solution of the $N_0=1$ small $\mu$ problem for $0 < m < N$.

The perspicacious reader will have noticed that the problem we are solving in this section does not reduce exactly to the problem we posed when we considered the pure-birth asymmetric model.  The equations are the same with the identification that $T_N(m)$ in our section maps to $P_N(m)/((N-m)r+m)$ in that section.  However, the probabilities we construct from the $T_N$ do not correspond to the $P$'s in the previous section.  Here, we weight each
state by the time spent in the state, which we did not do there.  To leading order in $N$, however, the time spent in a state is independent of the state (as long as $m\ll N$), so the large-$N$ solution is identical in the two ensembles.

Note that when $x=0$, we get
\begin{equation}
P(0|N) = \exp\left(-\mu N\frac{b_w}{rb_m}  \,{}_2F_1\!\left(1,r;1+r;\frac{ d_m}{b_m}\right)\right) ,
\end{equation}
which  is equivalent to a well-known result of Ref. \cite{iwasa}. The advantage of our approach is that it is clear what approximations have to be made to obtain this result, something that is not overly clear in this earlier reference.  It should also be noted in this context that the result obtained in Ref. \cite{iwasa} for $P(m|N)$ in the small $\mu N$ limit is not correct.

We can use Eq. (\ref{gensmallt}) to find the large $\mu N$, $m\sim \mu N$ limiting distribution, where the branch cut integral is as usual dominated by the singular point at $y=1$. Given that the analytic structure here, as we have seen,
is exactly the same as for the asymmetric pure-birth case, we immediately can conclude that the distribution for large $\mu N$ reproduces the one-sided
$\alpha$-stable distribution with index $r$.
The shift parameter is 
\begin{align}
\nu &= \mu N \frac{b_w}{b_m}\frac{b_m-d_m}{b_w-d_w} \frac{b_w-d_w}{b_w-d_w-b_m+d_m}\frac{b_m}{b_m-d_m} \nonumber\\
&= \mu N \frac{b_w}{b_w-d_w-b_m+d_m} .
\end{align}
That this is correct can be seen from the calculation of $\overline{m}$ in the original DL model:
\begin{align}
\overline{m} &= \int_0^T \mu b_w N_0 e^{(b_w-d_w)t} e^{(b_m-d_m)(T-t)} dt \nonumber\\
&= \mu (N-N_0) \frac{b_w}{b_w-d_m-b_m+d_m} .
\end{align}
For the scale parameter, we get
\begin{align}
c &= \frac{b_m}{b_m-d_m}\left[ \mu N \left(\frac{b_w}{b_m}\right) \left(\frac{b_m-d_m}{b_w-d_w}\right)\frac{b \pi}{\sin \pi b} \cos \frac{\pi b}{2} \right]^{1/b} \nonumber\\
&= \left(\frac{b_m}{b_m-d_m}\right)^{1-1/b} \left[ \mu N \left(\frac{b_w}{b_w-d_w}\right)\frac{b \pi}{2\sin\frac{ \pi b}{2}}  \right]^{1/b} .
\end{align}

Lastly, for completeness, we write down the results for the symmetric birth-death process.  Here $r=1$ and the large $\mu N$ limit is thus again a Landau distribution.  For $P(0|N)$ we have
\begin{equation}
P(0|N)\equiv = e^{-\mu N b_w/d_w \ln \frac{b_w}{b_w-d_w}} .
\end{equation}
For the large $\mu N$ limit, we have
\begin{equation}
P(m|N) \approx \frac{b-d}{\pi \mu N b} \int_0^\infty dt e^{-t m/(\mu N b/(b-d)  -  t \ln t + t\ln\mu N } \sin \pi t ,
\end{equation}
which is again a Landau distribution, $P_\textrm{Landau}(m(b-d)/\mu N b- \ln \mu N)$. Amazingly, introducing death to the LD problem hardly makes a change in the answer, just renormalizing the variable $m$ by the ratio $(b-d)/b$.

\section{Comparison with the fixed time ensemble}

\subsection{ Pure birth model}
The most general solution of LD problem is given in the recent paper by Antal and Krapivsky (AK)~\cite{antal}, who solve the problem of determining the probability distribution at a fixed time. We first present their solution, modifying their notation slightly so as not to interfere destructively with the notation we have been using until now. They write down the backward version of the master equation for the probabilities of $w$ wild-types and $m$ mutants in the form
\begin{align}
\frac{dP^w}{dt} (w,m,t)& =  b_w (1- \mu) P^{ww} (w,m,t) + \mu b_w  P^{wm}(w,m,t)\nonumber\\
&\qquad\qquad{} - (b_w + d_w) P^w (w,m,t)  +d_w \delta_{w,0} \delta_{{m,0}} ,\nonumber \\
\frac{dP^m}{dt} (w,m,t) & =  b_m P^{mm} (w,m,t) - (b_m + d_m) P^m (w,m,t) +d_m \delta_{N,0} \delta _{m,0} .
\end{align}
In this expression the division and death rates of the wild-type ($w$) and mutant ($m$) subpopulations are denoted by $b$ and $d$ with the proper subscript. The superscript denotes a process which starts from a single wild-type individual (superscript $w$), a single mutant individual (superscript $m$), two wild-type ($ww$) and one of each ($wm$). A great simplification occurs in passing to the generating function version, in that the independent time evolution of any two individuals leads to factorization of the two-individual probabilities, so that the generating function for $P^{ww}$ is the square of the generating function for $(P^w)$, with similar results for $P^{wm}$ and $P^{ww}$. This then leads to a pair of coupled ODE's for $W(x,y)$ and $M(x,y)$, the generating functions defined as
\begin{equation} W(x,y;t) \equiv \sum x^m y^w P^w (w,m;t);\qquad  M(x,y;t) \equiv \sum x^m y^w P^m (w,m;t) ,
\end{equation}
and the equations are
\begin{eqnarray} 
\frac{\partial W}{dt} & = & b_w (1 -\mu) W^2 + \mu b_w M W  + d_w  - (b_w +d_w) W , \nonumber \\
\frac{\partial M}{dt}  & = & b_m M^2 + d_m -(b_m+d_m)M .
\label{Aeq}
\end{eqnarray}

To see how the solution of this system works and how it relates to our approach, we return for the moment to the simpler case of no death and symmetric division $b_w=b_m \equiv b$, and recall that the AK derivation so far is for $N_0 =1$. We can rewrite these equations by picking the unit of time to set the wild-type growth rate $b(1-\mu)$ equal to  1. This leads to the new system
\begin{eqnarray} 
\frac{\partial W}{dt} & = & W^2 + \nu  M W   - (1+\nu) W \nonumber \\
\frac{\partial M}{dt}  & = & \lambda M^2   - \lambda M
\label{A-birtheq}
\end{eqnarray}
with $\nu\equiv\mu/(1-\mu)$, $\lambda\equiv1/(1-\mu)$. The initial condition for the $M$ equation is that $M(t=0) = x$, since by definition the state defining $M$ starts with one mutant cell. The solution of this equation is
\begin{equation} 
M = -\frac{z}{1-z};  \qquad\qquad z=-\frac{x}{1-x} e^{-\lambda t} \label{Z-DEF}\ . \end{equation} 
Substituting this into the $W$ equation, one obtains the solution
\begin{equation}
W=-\frac{Cz}{(1-z)^\mu + C - Cz}\ .
\end{equation}
The constant of integration $C$ is picked to satisfy the initial condition $W(t=0) =y$, yielding
\begin{equation}
C=-\frac{y(1-x)^{1-\mu}}{y-x}\ .
\end{equation}
From this, one can obtain directly the distribution of mutant numbers, independent of the number of wild-type, given by $F(x,t) \equiv W(x,y=1;t)$; setting $y=1$ gives $C=1$ and using the mean population size $\N \equiv e^{\lambda t}$, we obtain
\begin{equation}
F(x) = \frac{x}{x + \N(1-x)\left[1 - (1-x+x/\N)^\mu\right]} .
\end{equation}

\begin{figure}
 \includegraphics[width=0.7\textwidth]{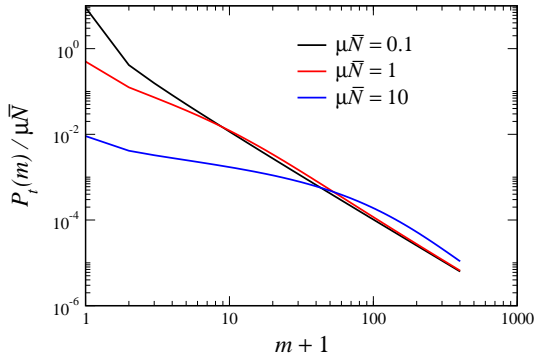}
 \caption{The scaled distribution of mutants, $P_t(m)/\mu\N$, in the AK fixed time ensemble, for times corresponding to $\mu\N =0.1$, 1 and $10$.}
 \label{figKrap}
 \end{figure}

To compare this result with our finding given above in Eq. \ref{lea-coulson}, we go to the scaling limit, which is small $\mu$, large $\N$, finite $m$ (so that the number of wild-type is approximately equal to the total population). We can then drop the $x/\N$ term, and we have
\begin{equation}
F(x) \approx \frac{x}{x - (1-x)\mu \N \ln (1-x)} ,
\label{Krapinner}
\end{equation}
so that indeed $\mu \N$ is the relevant control parameter.  This is {\em{not}} the Lea-Coulson generating function, as has already been noted in our abbreviated initial paper~\cite{pnas}.  In particular, $P_0 = 1/(1+\mu \N)$, so that it does not decay exponentially with $\mu \N$ as we found above.  The difference lies in the fact that the AK calculation is performed for {\em{fixed time}}, not at {\em fixed total population $N$} as before.
For large time, the distribution of the number of wild-type cells, independent of the number of mutants can be shown to be geometric:
\begin{equation}
P^w(w) \approx \frac{1}{\overbar{w}(t)} e^{-w/\overbar{w}(t)} ; \qquad\qquad \overbar{w}(t) = e^{b (1-\mu) t} .
\end{equation}
From this, given that the probability of zero mutants given $w$ wild-type cells is, for small $\mu$, $\exp(-\mu (w-1))$, it immediately follows that the probability of zero mutants is given by
\begin{equation}
P_0 \approx \int_0^\infty \frac{dw}{\overbar{w}(t)} e^{-\mu w} e^{-w/\overbar{w}(t)} \approx \frac{1}{1 + \mu\N},
\end{equation}
in agreement with the AK calculation. Clearly, then, the  configurations at  fixed time with  small numbers of wild-types and thus a relatively large probability of never having spawned a resistant cell increase $P_0$ drastically.  In Fig. \ref{figKrap} we present the distribution resulting from the AK fixed time ensemble, for times corresponding to $\mu \N=0.1$, 1 and $10$.  We see that in all cases the distribution is monotonic, and looks nothing like the Lea-Coulson result, nor does it go over to the Landau distribution at large $\mu \N$.
 
  Looking at $F(x)$ for small $\mu \N$, we see that it does reproduce the small $\mu N$ limit of the Lea-Coulson answer.  More generally,  the structure of the singularity at $x=1$ is the same of for Lea-Coulson, so that fixed-time
  ensemble does show the same $1/m^2$ behavior as our fixed-$N$ ensemble for the intermediate asymptotic regime $1 \ll m\ll N$.  The behavior in the 
  cut-off regime $m \sim N$ is not surprisingly very different from the fixed-$N$ behavior, since the shift of the pole of $F(x)$ to $x=1/(1+1/\N)$ means that the decay of $P(m)$ is exponential for $m \sim N$

Of course, the full AK solution contains information about the wild-type distribution encoded in its dependence on $y$ which has been dropped in the derivation of $F$. If we return to the full solution for $W$, we can of course restrict ourself to a fixed value of $m=\N$ by finding the coefficient of the term that goes like $y^N$. This is still not exactly the same problem we solved.  Our initial  ensemble allows the elapsed time to vary, whereas this new result fixes both the time and the total population. However, the distribution of mutants as we have seen is controlled by the total population, and we do not expect the time taken to reach this population to matter significantly, at least as long as it is close to the average time.  Further, as we have already noted, for $m\ll N$, the total population $N$ and the wild-type population $w$ are the same to leading order in $N$.  Thus, undaunted, we continue. The projection of $W(x,y)$ onto fixed $m=\N$, which we denote $W_\N(x)$ can be done exactly, since the $y$ dependence of $W$ is simple and takes the  form
\begin{equation}
W = A_1 \, y/(A_2 \, y + A_3)
\end{equation}
 with
\begin{equation}
 A_1 = z,\qquad  A_2 = z-1 + \frac{[(1-z)(1-x)]^\mu}{1-x}, \qquad  A_3=-\frac{x[(1-z)(1-x)]^\mu}{1-x},
 \end{equation}
 and where as before $z=-\frac{x/\N}{1-x}$. 
 Thus,
 \begin{equation}
 W_\N(x) = {\cal{N}}\frac{A_1}{A_3} \left(-\frac{A_2}{A_3}\right)^{\N-1} .
\end{equation}
where ${\cal{N}}$ is a normalization factor, chosen so that $W_\N(1)=1$.  The non-trivial normalization factor arises because we do not know a priori the overall probability of finding exactly the correct value of $N$ in the overall ensemble, and hence this must be divided out to obtain a meaningful probability distribution. Now, in the limit $x\to 1$, 
\begin{align}
z \to -\frac{1}{\N(1-y)} \to -\infty , &\qquad A_2 \to -\frac{1}{\N(1-y)} + \frac{\N^{-\mu}}{1-y}  = \frac{\N^{-\mu}(1-\N^{\mu-1})}{1-y},\nonumber\\
&  A_3  \to -\frac{\N^{-\mu}}{1-y} .
\end{align}
Thus, we can find the value of the normalization constant
\begin{equation}
{\cal{N}} = \N^{1-\mu} \left(1- \N^{\mu-1}\right)^{1-\N} \approx \N^{1-\mu} e^{1+\mu \ln \N}  =\N e^1 .
\end{equation}
Given this, we can now calculate the leading large-$\N$ behavior of  $W_{\N}(x)$,
which we wish to compare to $F(x)$.  For fixed $x$ and large $\N$, $z\to 0$ and 
\begin{equation}
A_2/A_3 = \frac{[(1-z)(1-x)]^{1-\mu}}{x} - \frac{1}{x} \approx-1 + \frac{1}{\N} - \mu\frac{(1-x)\ln(1-x)}{x} .
\end{equation}
Also,
\begin{equation}
A_1/A_3 =- \frac{z(1-x)}{x[(1-z)(1-x)]^\mu}  \approx \frac{1}{N} .
\end{equation}. Finally we obtain 
\begin{equation}
W_\N(x) \approx Ne^{1} \left(\frac{1}{N}\right) e^{-1+ \mu N\frac{(1-x)\ln(1-x)}{x}} = (1-x)^{\mu N(1-x)/x} .
\end{equation}
This is just the Lea-Coulson result that we had obtained above. Thus, our expectations that post-projecting the fixed-time ensemble for fixed wild-type abundance would reproduce the fixed population ensemble result in the intermediate asymptotic regime are indeed realized.

The above calculation was for $N_0=1$.  For general $N_0$, the AK result for the generating function $W(x,y;N_0)$ is immediate:
\begin{equation}
W(x,y;N_0)= [W(x,y)]^{N_0} ,
\end{equation}
due to the independence of the life histories of the initial $N_0$ wild-type.  Clearly then, the AK result, i.e. conditioning on fixed time, is $N_0$ dependent,
as opposed to the $N_0$ independence of the fixed total population result, for $m \ll N$.  We can generalize our calculation of the fixed $m=\N$ projection  of the AK result to general $N_0 \ll \N$.  Now, we have $\exp(\lambda t)=\N/N_0$. Then, performing the Taylor expansion of $W^{N_0}$ we get
 \begin{equation}
 W_\N(x;N_0) = {\cal{N}}\left(\frac{A_1}{A_3}\right)^{N_0} {\N - 1\choose N_0-1}  \left(-\frac{A_2}{A_3}\right)^{\N-N_0} .
\end{equation}
The normalization factor is now
\begin{align}
{\cal{N}} &= \left(\frac{\N}{N_0}\right)^{N_0(1-\mu)}\N^{1-N_0}(N_0-1)! \left[1- \left(\frac{\N}{N_0}\right)^{\mu-1}\right]^{N_0-\N}\nonumber\\
& \approx \N\left(\frac{e}{N_0}\right)^{N_0}(N_0-1)! .
\end{align}
Then,
\begin{align}
W_{\N}(x;N_0) &= {\cal{N}}\left( \frac{N_0}{\N}\right)^{N_0} {\N - 1\choose N_0-1} \left(1-\frac{N_0}{\N}+\mu\frac{(1-x)\ln(1-x)}{x}\right)^{\N-N_0}\nonumber\\
&\approx(1-x)^{\mu \N(1-x)/x} ,
\end{align}
recovering, as we expect, the $N_0$-independent Lea-Coulson result.  It is remarkable that the fixed population ensemble results are much more similar to
the original semi-deterministic model of Luria and Delbr\"uck, with its universal aspects such as $N_0$ independence and large $\mu N$ $\alpha$-stable distribution than is the fixed time ensemble.

\subsection{General Asymmetric Birth-Death Model}
We chose to first present our projection method in the simplest possible context of the pure-birth, symmetric model, as in this case the AK solution is expressible in terms of elementary functions and the structure of the calculation is most transparent.  The procedure, however, works in essentially the same way for the most general asymmetric birth-death model for which AK presented their solution for the fixed time ensemble.  The model is now
parameterized in terms of wild-type and mutant birth rates, $b_w$, $b_m$ as well as wild-type and mutant death rates, $d_w$, $d_m$.  AK choose to scale time so that the non-mutating wild-type birth process $W\to WW$ has rate unity.  AK then introduce the following rates:
\begin{align}
\alpha_m \equiv \frac{b_m}{b_w(1-\mu)}, &\qquad \beta_w \equiv \frac{d_w}{b_w(1-\mu)}, \qquad \beta_m  \equiv \frac{d_m}{b_w(1-\mu)}, \qquad \nu = \frac{\mu}{1-\mu}, \nonumber\\
& \lambda_w=1 - \beta_w - \nu, \qquad \lambda_m = \alpha_m - \beta_m .
\end{align}
AK also introduce the auxiliary parameter
\begin{equation}
\omega \equiv -\frac{\lambda_w}{2} - \frac{\nu\beta_m}{2\alpha_m} + \sqrt{\left(\frac{\lambda_w}{2}+\frac{\nu\beta_m}{2\alpha_m}\right)^2 + \frac{\nu\lambda_m}{\alpha_m}} .
\end{equation} 
Note that in the small $\mu$ limit, $\omega$ is of order $\mu$:
\begin{equation}
\omega \approx \mu \frac{b_m-d_m}{b_w-d_w}\frac{b_w}{b_m}.
\end{equation}
AK define as well 
\begin{align}
a\equiv \frac{\omega}{\lambda_m} \approx \frac{\mu b_w^2}{b_m(b_w - d_w)}, &\qquad b \equiv \frac{\omega + 1 - \beta_w}{\lambda_m} \approx \frac{b_w - d_w}{b_m - d_m},\nonumber\\
&  c \equiv 1 + a + b - \frac{\nu}{\alpha_m}\approx b + 1 ,
\end{align}
where the approximations are correct to leading order in $\mu$.
Then, the full solution derived in AK takes the extremely complicated form
\begin{equation}
W(x,y;t) = 1+ \omega + \frac{T_1 y + T_2}{T_3 y + T_4} = 1+ \omega + \frac{(T_1 y + T_2)}{T_4} \sum_k (-1)^k \left(y\frac{T_3}{T_4}\right)^k ,
\end{equation}
where
\begin{align}
T_1 &= z^{c+1} z_0 F_3(z) F_2(z_0) + (c-1) z_0^c F_2(z) F_1(z_0) z - z^2 z_0^cF_4(z) F_1(z_0) ,\nonumber\\
T_2 &= z^{c+1} z_0 F_3(z) \left[F_2(z_0)(\lambda_m(c-1)-1-\omega) - z_0F_4(z_0)\lambda_m \right] \nonumber\\
&\qquad{}- z_0^c z(c-1)F_2(z)\left[F_1(z_0)(1+\omega) + \lambda_m F_3(z_0)z_0\right] \nonumber\\
&\  {} \qquad + z_0^c z^2 F_4(z) \left[ F_1(z_0)(1+\omega) + \lambda_m z_0 F_3(z_0)\right] ,\nonumber \\
T_3 &= z^c z_0 F_1(z) F_2(z_0) - z_0^c z F_2(z) F_1(z_0) , \nonumber\\
T_4 &= z^c z_0 F_1(z) \left[F_2(z_0)((c-1)\lambda_m - 1 - \omega ) - z_0 \lambda_m F_4(z_0)\right]\nonumber\\
&\qquad{}
+ z z_0^c F_2(z) \left[F_1(z_0)(1+\omega) + \lambda_m z_0 F_3(z_0)\right] ,
\end{align}
with the function definitions
\begin{eqnarray}
F_1 &=& {}_2F_1(a,b;c;z) , \nonumber\\
F_2 &=& {}_2F_1(1+a-c,1+b-c;2-c;z) , \nonumber\\
F_3 &=& \frac{ab}{c} {}_2F_1(1+a,1+b;1+c;z)  , \nonumber\\
F_4 &=& \frac{(1+a-c)(1+b-c)}{2-c}{}_2F_1(2+a-c,2+b-c;3-c;z) ,
\end{eqnarray}
and
\begin{equation}
z=\frac{1}{\M}\left[1 - \frac{\lambda_m}{\alpha_m(1-x)}\right], \qquad z_0 = \M z .
\end{equation}
Here, $\M\equiv e^{\lambda t}$ and is related to
 $\N$, the average number of wild-type cells, by
\begin{equation}
\N \equiv e^{\lambda_w t} = \M^{\lambda_w/\lambda_m} \approx \M^b .
\end{equation}
Then, $W_{\N}(x)$ is given by
\begin{equation}
W_\N(x) = {\cal{N}}\frac{T_1 T_4 - T_2 T_3}{T_4^2} \left(-\frac{T_3}{T_4}\right)^{N-1} .
\end{equation}

For the calculation of the $T_1T_4-T_2T_3$ factor, it is sufficient to work to zeroth order in $\mu$.  To this order $F_1=F_2=1$ and $F_3=F_4=0$.  Then, we have
\begin{align}
T_1 T_4 - T_2 T_3 &\approx \left[ bz_0^c z \right]\left[ z^c z_0 (b\lambda_m - 1) + z z_0^c\right] - \left[-bz_0^c z \right]\left[z^cz_0-z_0^c z\right] \nonumber\\
&= b^2\lambda_m z_0^{c+1} z^{c+1} .
\end{align}
For the $T_4^2$ term in the denominator, things are even simpler since we can use the fact that $z\ll z_0$:
\begin{equation}
T_4^2 \approx z_0^{2c} z^{2} ,
\end{equation}
so that 
\begin{equation}
\frac{T_1 T_4 - T_2 T_3}{T_4^2} \approx b^2 \lambda_m \left(\frac{z}{z_0}\right)^b \approx b\lambda_m/N .
\end{equation}

For the calculation of $T_3/T_4$ we have to keep both the first $1/N$ and $\mu$ corrections since the ratio is to be raised to the $\N$th power.
We have, since $a \sim {\cal{O}}(\mu)$
\begin{align}
F_1(x) &\approx 1 + a \sum_{n=1}^\infty \frac{\Gamma(b+n)\Gamma(b+1)}{\Gamma(b)\Gamma(b+1+n)n} x^n \nonumber\\
&= 1 + \frac{a}{b}\left[1-b\ln(1-x)-{}_2F_1(1,b;1+b;x)\right] \equiv 1 + \mu F_1^1(x) ,
\end{align}
and
\begin{align}
F_3(x) &\approx  a \frac{\Gamma(b+1)}{\Gamma(b)} \sum_{n=1}^\infty \frac{n\Gamma(n)\Gamma(b+n)}{n!\Gamma(b+n+1)}x^{n-1} \nonumber\\
&= a \frac{{}_2F_1(1,b;b+1;x)-1}{x} \equiv \mu F_3^1(x) ,
\end{align}
and so
\begin{equation}
T_3 \approx -z_0^cz \left[1 + \mu F_1^1(z_0) - \frac{1}{\N}\right] .
\end{equation}
Similarly,
\begin{equation}
T_4\approx z_0^c z\left[ 1 + \omega + \mu \lambda_m z_0 F_3^1(z_0) + \mu F_1^1(z_0) + \frac{b\lambda_m-1)}{\N}\right] .
\end{equation}
Thus
\begin{equation}
\frac{T_3}{T_4} \approx -\left[ 1 - \frac{b\lambda_m}{\N} - \omega - \mu\lambda_m z_0 F_3^1(z_0) \right] .
\end{equation} 

Thus, we have our result
\begin{equation}
W_\N(x) \approx {\cal{N}}\left(\frac{b\lambda_m}{\N}\right)\exp\left(-b\lambda_m - \N\omega - \mu \N \lambda_m z_0 F_3^1(z_0)\right) .
\end{equation}
Since, when $x\to 1$, $z_0 \to -\infty$ and $\mu F_3^1(z_0) \to -a/z_0$, the condition $W_\N(1)=1$ implies
\begin{align}
W_\N(x) &\approx \exp\left(- \mu \N \lambda_m z_0 (F_3^1(z_0)+a/z_0)\right) \nonumber\\
&= \exp\left(-\mu N\left(\frac{b_w}{b_m}\right) \left(\frac{b_m-d_m}{b_w-d_w}\right){}_2F_{1}(1,b;b+1;z_0)\right) .
\end{align}
This recovers the result we obtained above, as expected.

It is amazing how our relatively simple scaling distribution is encoded in the immensely more complicated AK formalism. Being exact, their equation must contain all the details regarding the outer region (when $m$ is the same order as $N$), must allow for arbitrary values of $\mu$ and $N$, and must reflect the intricate interplay of time and population size. Our approach is much less general, but has the distinct advantage of leading to answers that are simple enough for us to gain some insight into what is going on. In the last part of this section, we will see that we can employ a trick to recover our results from the AK formalism
in a much simpler fashion.

\subsection{An alternate method}

There is an alternate way to extract our desired fixed population distribution from the AK results.  We have seen that in the regime of interest,
$m \ll N$, the distribution is independent of $N_0$.  If we take $N_0$ large however, the wild-type dynamics becomes deterministic and so the fixed
time and fixed population ensembles become identical.  Thus, all we need to do is to take the AK results for large $N_0$, and this gives the desired
generating function directly.
We now proceed to carry this out for the general non-symmetric case.

Following the notation in AK, we focus on 
\begin{equation} W = 1 +\omega + \lambda _m \Psi (z) ,\end{equation}
with the definition
\begin{equation}
\Psi (z)  \equiv \ \frac{z^g F_3 (z) +C(1-g) F_2(z) +Cz F_4(z) }{ z^{g-1} F_1(z) +C F_2 (z)} ,
\end{equation}
where the $F's$ are the same as given above, with the constants taking the values (valid to leading order in $\mu$):
\begin{align}
\omega \approx \mu \frac{b_m-d_m}{b_w-d_w}\frac{b_w}{b_m};   &\qquad \lambda_m \approx  \frac{b_m-d_m}{b_w}; \nonumber\\
e\approx \mu \frac{b_w^2}{b_m(b_w-d_w)}; &\qquad  f\approx \frac{b_w-d_w}{b_m-d_m};\qquad g \approx d+1,
\end{align}
and where the constant $C$ is determined by the normalization condition $W(t=0) = y$:
\begin{equation}
C \ = \  z_0^g \frac  {\kappa F_1(z_0) - F_3(z_0)}{(1-f-\kappa z_0) F_2 (z_0) +z_0 F_4 (z_0)} ,
\end{equation}
with $z_0$ the value of $z$ at $t=0$ . Note that in general $z$ is proportional to $e^{\lambda _m t}$. Finally $\kappa \equiv -\frac{\omega}{\lambda_m z_0} \approx - \frac{e}{z_0}$.
This seems to be an completely unwieldy mess. But, we get an enormous simplification if we recall that we are interested in small $\mu$ and therefore can use the leading order results $F_1 =1$, $F_2 = 1$, and $F_4$ is $O(\mu )$ and hence can be dropped in the denominator, since $C$ is also $O(\mu )$. with the definition $F_3^1(z) = F_3^1(z)/\mu$, we can derive the much simpler leading order expression
\begin{equation}
C \approx z_0^g \frac{e/z_0 + \mu F_3^1(z_0)}{f} \equiv \mu C_1  .
\end{equation}

Finally, we use the fact that $z$ is small as it is related $e^{- \lambda _m t}$ which equals  $N_0/N$ raised to the positive power $f$.
Thus, $W(x)$ is given by  
\begin{equation}
W(x) =\left(1 + \lambda_m \Psi(z)\right)^{N_0} =   \exp\left(-\mu N\left(\frac{b_m-d_m}{b_w}\right) \left(d/\mu+z_0F_3^1(z_0)\right)\right) .
\end{equation}
We need an explicit expression for $F_3^1(x)$.  We have
\begin{align}
F_3^1(x)&=  \frac{a}{\mu} \frac{\Gamma(b+1)}{\Gamma(b)} \sum_{n=1}^\infty \frac{n\Gamma(n)\Gamma(b+n)}{n!\Gamma(b+n+1)}x^{n-1} \nonumber\\
&= \frac{ab}{\mu} \sum_{n=1} \frac{x^{n-1}}{b+n} \nonumber\\
&= \frac{a}{\mu} \frac{{}_2F_1(1,b;b+1;x)-1}{x} .
\end{align}
  Thus, we have the final expression
\begin{align}
W(x) &=   \exp\left(-\mu N\left(\frac{b_m-d_m}{b_w}\right) \left(\frac{b_w^2}{b_m(b_w-d_w)}\right){}_2F_{1}(1,b;b+1;z_0)\right)\nonumber\\
&= \exp\left(-\mu N\left(\frac{b_w}{b_m}\right) \left(\frac{b_m-d_m}{b_w-d_w}\right){}_2F_{1}(1,b;b+1;z_0)\right) ,
\end{align}
again reproducing what we have previously calculated.

\section{Conclusions}
We have analyzed here the pure birth as well as the birth-death stochastic Luria-Delbr\"uck model in the limit of small mutation rate $\mu$ and
large current population, $N$, where the product $\mu N$ is unrestricted.  The answer in all cases exhibits  an intermediate asymptotic power-law decay $m^{1+r}$ for $1 \ll m\ll N$, where $r$ is the ratio of the wild-type net growth rate to the mutant net growth rate.  For small $\mu N$, the distribution is monotonically decreasing, and approaches the pure power-law already for $m$'s of order unity.  As $\mu N$ increases the probability of a small number of mutants decreases sharply, and above a critical $\mu N$ of order 1, the distribution develops a peak.  The power-law decay sets in then only a distance past this peak.  For large $\mu N$, the distribution becomes the one-sided L\'evy distribution with index $r$.  The distribution for $m \ll N$ is in all cases independent of the  initial wide type colony size $N_0$.  It is also independent of the precise nature of the ensemble, as long as there are sufficient numbers of wild-type individuals in all members of the ensemble with non-vanishingly small weighting. This can be guaranteed by fixing $N = \bar{N}$ at either fixed time or in a time average, or alternatively by having a large number for the number $N_0$ of initial wild-type individuals. The power-law tail, as well as the strong dependence of the distribution on $N_0$ for $m \sim N$ has significant implications for tasks such as using measured data to infer mutation rate \cite{kepler,gerrish}. This also has implications for the variation expected in patient responses to administered therapy, although it seems that the {\em extrinsic} variation due to different rates dominates over the {\em intrinsic} variation due to process stochasticity ~\cite{cancer-research}.

\begin{acknowledgements}
This work was supported by the NSF Center for Theoretical Biological Physics, (grant no. PHY-1308264). HL was also supported by CPRIT Scholar program of the State of Texas, and DK was also supported by the Israeli Science Foundation. \end{acknowledgements}

\end{document}